# Factors Influencing mHealth Acceptance among Elderly People in Bangladesh

**Md. Rakibul Hoque**
Department of Management Information Systems
University of Dhaka
Dhaka, Bangladesh
Email: rakibul@du.ac.bd

**Dr. Golam Sorwar**
School of Business and Tourism
Southern Cross University
East Lismore NSW 2480, Australia
Email: Golam.Sorwar@scu.edu.au

## Abstract

mHealth (mobile health) services are becoming increasingly important form of information and communication technology (ICT) enabled delivery for healthcare. Given widespread mobile adoption and potential benefits mHealth can bring, it is important to understand what motivates users' acceptance and adoption of its applications. However, little research has been done to draw any systematic study of potential users' adoption and usage of mHealth technology. The aim of this study is to determine factors influencing the adoption and use of mHealth technology and services by elderly in Bangladesh. In this study, a theoretical model based on the Unified Theory of Acceptance and Use of Technology (UTAUT) will be developed and tested to determine the elderly behavioural intention to adopt mHealth application. The study is expected to identify and validate relationship between attitudinal constructs presented in the UTAUT and acceptance of mHealth services in Bangladesh. The study will also investigate the moderating role of gender, experience with mobile phone use and their influence on the adoption of mHealth technology and services.

**Keywords:** mHealth adoption, Elderly, UTAUT.

## 1 INTRODUCTION

Over the past decade, health care sectors have experienced tremendous changes in most of the countries in the world due to a rapid advancement in information and communication technology (ICT). Recent evidence suggests that mHealth (mobile health) be the blessing of ICT and is probably one of the most prominent services with noticeable effect on the development of healthcare sector (Nessa et al., 2008; Sharifi et al., 2013; Hoque et al., 2014). The Global Observatory for e-Health (Kay, 2011) defined mHealth as "medical and public health practice supported by mobile devices, such as mobile phones, patient monitoring devices, personal digital assistants (PDAs), and other wireless devices." mHealth plays critical role in providing better access to healthcare facilities equally to patients, physician, nurses, and other healthcare personnel, to increase care quality and improving collaboration (Khalifehsoltani and Gerami, 2010). It has the ability to interact with the users with greater flexibility and frequency without being constrained by time and/or location (Free et al, 2010). It enables users to access personalized and interactive health services due to their mobility, portability and ubiquity (Akter et al, 2010). Studies have shown that healthcare services via mobile phone appear significant for elderly patients who are in need of most healthcare demand (Pew Internet, 2010).

Globally, the proportion of older population is increasing faster than any other age group due to an increased life expectancy and a decrease in birth rates (AIHW, 2012). In 2013, there were about 840 million people (10% of overall population) over the world aged 60 years or more (WHO, 2013). By 2025 and 2050, this figure is expected to reach to around 1.2 billion and 1.9 billion respectively (UN, 2012). In developing countries, where 80% of older population live, the proportion of the elderly is expected to increase by 12% by 2025 (Hutton, 2008).

Bangladesh is one of the twenty countries in the world with the largest elderly populations (Munsur et al., 2013). By 2025, it, along with China, India, Indonesia and Pakistan, will account for about 44% of the world's total elderly population (Chaklader et al, 2003). The most recent census (population census, 2011) data shows that 7.4% of total population in Bangladesh is 60 years or more (BBS, 2011). The proportion of elderly population in Bangladesh is projected to increase to 11.9% and 17% by 2035 and 2050 (Desa, 2009). It is obvious that elderly populations are more susceptible to chronic diseases, physical disabilities and mental incapacities (Kabir, 2001). A study indicates that more than 95% of elderly in Bangladesh experienced health problems with most of them reported multiple health issues (Kabir et al, 2003). Chronic diseases are the leading healthcare cost related to hospitalization and re-hospitalization (Bardhan et al., 2012). Mobile phone users are also rapidly increasing in Bangladesh. According to Bangladesh Telecommunication Regulatory Commission (BTRC), the number of Mobile Phone users in Bangladesh was 125.971 million at the end of May 2015, with an annual growth of 7.85% (BTRC, 2015). This data, therefore, implies the potentiality of various mHealth initiatives and interventions in Bangladesh (Ghorai et al., 2013). Through mobile health services, elderly people can seek medical advice, register for or check appointments, access medical test results, and seek post-diagnostic treatment for active prevention at their convenience (Deng et al, 2014).

Given the potential benefits mHealth can bring to older people, it is important to understand the users' (elderly population) intention to use the technology. Usually, elderly people lag behind in technology-based innovation and encounter difficulties and challenges to use it (Beer and Takayama, 2011). Research found that elderly people have less technological self-efficiency, higher level of anxiety, less control over ICT, fewer new technology skills that are likely to negatively affect acceptance of innovative technology such as mHealth (Steele et al., 2009). In addition, physical and cognitive capabilities possibly also reduce their intention to use innovative technology (Czaja et al., 2006).

In Bangladesh, mHealth service is still in its infancy stage and requires a user adoption process, especially for elderly users with unique physical and psychophysical characteristics (Ahmed et al., 2014; Akter and Ray, 2010). It is, therefore, timely to study elderly's mHealth adoption and acceptance behavior for the development of appropriate mobile health services in Bangladesh. However, little research has been done to draw any systematic study of elderly's adoption and usage of mHealth. A study by Akter et al. (2010) investigated the mHealth adoption and acceptance factors in Bangladesh from service provider perspective. This study, however, has little focus on mHealth adoption from users' perspective including elderly. Based on the above literature review, the following research question is formulated in this study.

- What are the key factors influencing elderly acceptance and adoption of mHealth services in Bangladesh?

The research objectives pursued in order to answer the above research question are to:
- determine key factors influencing the adoption and use of mHealth technology and services by the elderly in Bangladesh.
- determine the moderating roles of gender on the adoption of mHealth.
- determine the moderating roles of user's experience with mobile phone use on the adoption of mHealth.

The Unified Theory of Acceptance and Use of Technology (UTAUT) will be used, as the theoretical underpinning of the research, to understand the factors that influence the elderly acceptance and adoption of mHealth services in Bangladesh.

The organization of the rest of the paper is as follows. A brief review of the relevant literature is presented in Section 2. We discuss the development of theoretical framework and hypotheses in Section 3 which is followed by a detailed description of methodology in Section 4. Finally we conclude the paper in Section 5.

## 2 LITERATURE REVIEW

There has been moderate number of literature available on mHealth programs around the world. This section provides a brief review of existing literatures. Mobile tools have been utilized by community health workers for a broad range of healthcare services, specifically for maternal and child health and HIV/AIDS (Braun et al, 2013). It is quite common that community health workers collect field-based health data, receive alerts and reminders, and facilitate health education sessions (Källander et al., 2013). Sieverdes (2013) reported that mobile health technology provides guidelines related to self-monitoring of blood-glucose, medication management, medical nutrition therapy, physical activity and resistance training, and diabetes self-management. Pharow and Blobel (2008) analyzed some specific aspects of IT like mobile technology towards mobile health which aims to increase citizens and patients' consciousness, assurance, and acceptance in health care. Lefebvre (2009) provided a brief background on cell phone and mobile technology use in public health.

According to Lai (2005), ICT, such as mobile phone, paves the way for social and economic development. It saves time and money by disseminating information and advices of experts. Free et al. (2013) argued that mobile technology is an influential gadget rendering individual level assistance to health care consumers. Hameed (2003) provided a discussion on the application of mobile technology in the UK's health care service context and pointed out that some notable benefits can be derived from the integration of it into prevailing information infrastructure. Zhang et al. (2010) incorporated a partial least square modelling data analysis which revealed that the main factor in the adoption of mobile health is the perception of users' in the usefulness of that technology. Adibi (2012) demonstrated that transmitting health-related information through mobile applications based on 3G and 4G networks are emerging over time. The author mentioned about the logical and practical solution for mHealth through BlackBerry Enterprise Server.

Lim et al. (2011) revealed that mobile phones is found to be complementary to online health information and considered as an alternative source to the internet for seeking health information in Singapore. According to Robertson (2010), mobile phones are used extensively, though the internet access is limited in Sri Lanka, to gather health related information. Poropatich et al. (2013) highlight the applications of mobile communications by the US Army through the various levels of care in overseas locations where health services are more challenging. According to Vita wave consulting (2010), mHealth can effectively combat healthcare challenges by lowering costs, facilitating wide access and suitable solutions. ITU (2010) expected that mHealth would bring a widespread transformation in healthcare by improving care of patients and custom-made health services. A study conducted by UN (2010) revealed that 51 extensive mHealth programs have been already implemented in 26 developing countries around the world.

While substantial research has been done on mHealth in developed countries, very little and insufficient empirical study has been conducted in developing countries, especially in Bangladesh. Ahmed et al. (2014) pointed out some pragmatic issues regarding mobile phone-based health services provided by Directorate General of Health Services (DGHS) of Bangladesh. The study argued that this service does not have significant impact on the betterment of health service, however, generally the users held positive impression of the effectiveness of the service. Ashraf et al. (2010) argued that Bangladesh lagged behind in health sector. The majority of its population lives under the poverty line. It is, therefore, difficult for them to avail costly treatment. The above literature review shows that no research has been conducted yet to study the adoption of mHealth among elderly in the context of Bangladesh. In addition, no studies have been conducted to adopt the UTAUT Model in the arena of mHealth in Bangladesh.

## 3   THEORETICAL FRAMEWORK AND HYPOTEHSES

In order to analyse the research questions presented in introduction section, this study adopts a most influential user acceptance and behaviours analysis model, the Unified Theory of Acceptance and Use of Technology (UTAUT) (Venkatesh et al., 2003). Venkatesh et al (2003) claimed that the UTAUT can explain as much as 70% of the variance in intention. The objective of the UTAUT was to achieve a unified view of user acceptance (Venkatesh et al., 2003). It has been widely applied in different areas such as online bulletin boards (Marchewka et al., 2007), instant messengers (Lin & Anol, 2008), Web-based learning (Chiu & Wang, 2008), e-Health (Nuq, 2009), etc. The conceptual research model (Figure 1) developed based on the UTAUT model postulates that seven constructs: Performance Expectancy (PE), Effort Expectancy (EE), Social Influence (SI), Facilitating Condition (FC), Hedonic Motivation (HM), Price Value (PV), and Habit (HT), act as determinants of behavioural intentions and use behaviour. It is intended to determine if there is any significant relationship between the attitudinal constructs presented in the UTAUT model acceptance of mHealth services in Bangladesh.

In addition, UTAUT also posits the role of four key moderator variables: gender, age, experience, and voluntariness of use. Regarding the moderating effects, both age and voluntariness of use lie out of the scope of this research due to their irrelevancy with the purpose of the study. Previous study shows that gender and experience may have a considerable influence on users' acceptance of mHealth (Wang et al., 2003). This study will investigate the moderating role of gender and experience and their influence on the adoption of mHealth technology and services.

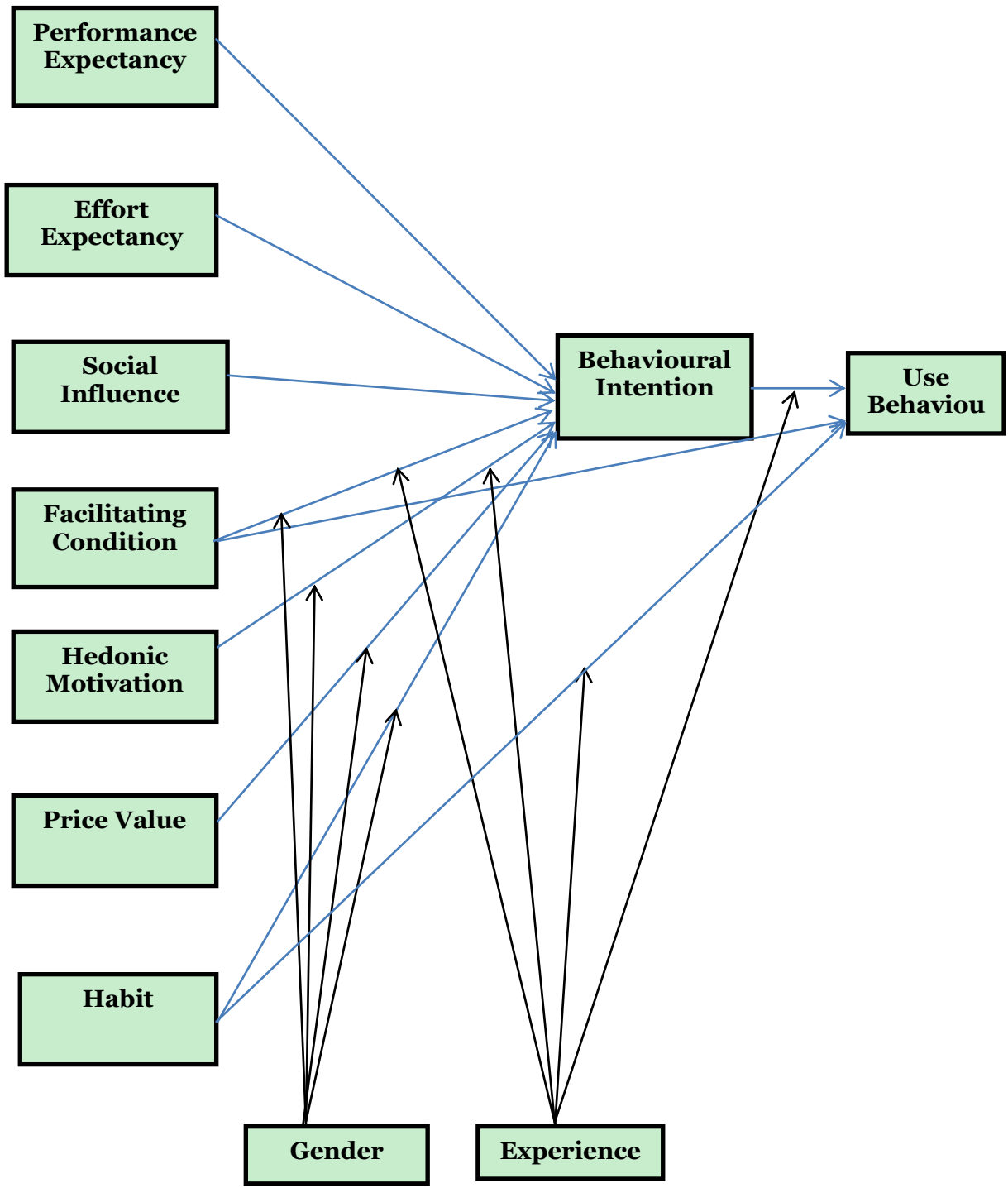

*Figure 1 Research Model (UTAUT2)*

On the basis of theoretical framework, this study proposes the following hypotheses.

*H1: Performance expectancy will have positive impact on elderly intention to use mHealth.*

*H2: Effort expectation will have positive impact on elderly intention to use mHealth.*

*H3: Social influence will have positive impact on elderly intention to use mHealth.*

*H4: Facilitating conditions will have positive impact on elderly intention to use mHealth.*

*H5: Facilitating conditions will have positive impact on elderly behavior of using mHealth.*

*H6: Hedonic motivation will have positive impact on elderly intention to use mHealth.*

*H7: Price value will have positive impact on elderly intention to use mHealth.*

*H8: Habit will have positive impact on elderly intention to use mHealth.*

*H9: Habit will have positive impact on elderly behaviour of using mHealth.*

*H10: Behavioural intention will have positive impact on elderly behaviour of using mHealth.*

*H11: The influence of Facilitating conditions on elderly intention to use mHealth will be moderated by gender.*

*H12: The influence of Facilitating conditions on elderly intention to use mHealth will be moderated by experience.*

*H13: The influence of Hedonic motivation on elderly intention to use mHealth will be moderated by gender.*

*H14: The influence of Hedonic motivation on elderly intention to use mHealth will be moderated by experience.*

*H15: The influence of Price value on elderly intention to use mHealth will be moderated by gender.*

*H16: The influence of Habit on elderly intention to use mHealth will be moderated by gender.*

*H17: The influence of Habit on elderly intention to use mHealth will be moderated by experience.*

*H18: The influence of Behavioural intention on elderly behaviour of using mHealth will be moderated by experience.*

## 4 METHODOLOGY

### 4.1 Measures

All measures, shown in Table 1, for latent constructs presented in the proposed model are developed from prior studies and modified on the mHealth context in Bangladesh.

| Constructs | Items | Sources |
|---|---|---|
| **Performance Expectancy (PE)** | PE1. I find mobile health service useful in my daily life. | Kijsanayotin et al. (2009), Vankatesh et al. (2012) |
| | PE2. Using mobile health service helps me accomplish things more quickly. | |
| | PE3. Using mobile health service increases my productivity. | |
| **Effort Expectancy (EE)** | EE1. Learning how to use mobile health service is easy for me. | Vankatesh et al. (2012) |
| | EE2. My interaction with mobile health service is clear and understandable. | |
| | EE3. I find mobile health service easy to use. | |
| | EE4. It is easy for me to become skilful at using mobile health service. | |
| **Social Influence (SI)** | SI1. People who are important to me think that I should use mobile health service. | Kijsanayotin et al. (2009) |
| | SI2. People who influence my behaviour think that I should use mobile health service. | |
| | SI3. People whose opinions that I value prefer that I use mobile health service | |
| **Facilitating Condition (FC)** | FC1. I have the resources necessary to use mobile health service. | Ifinedo (2012), Vankatesh et al. (2012) |
| | FC2. I have the knowledge necessary to use mobile health service. | |
| | FC3. Mobile health is compatible with other technologies I use. | |
| | FC4. I can get help from others when I have difficulties using mobile health. | |
| **Hedonic Motivation (HM)** | HM1. Using mobile health service is fun. | Vankatesh et al. (2012) |
| | HM2. Using mobile health service is enjoyable. | |
| | HM3. Using mobile health service is very entertaining. | |
| **Price Value (PV)** | PV1. Mobile health service is reasonably priced. | Vankatesh et al. (2012) |
| | PV2. Mobile health service is a good value for the money. | |
| | PV3. At the current price, mobile health service provides a good value. | |
| **Habit (HT)** | HT1. The use of mobile health service has become a habit for me. | Vankatesh et al. (2012) |
| | HT2. I am addicted to using mobile health service. | |
| | HT3. I must use mobile health service. | |
| **Behavioural Intention (BI)** | BI1. I intend to continue using mobile health service in the future. | Vankatesh et al. (2012) |
| | BI2. I will always try to use mobile health service in my daily life. | |
| | BI3. I plan to continue to use mobile health service frequently. | |
| **Use Behavior (UB)** | UB1: Mobile health service is a pleasant experience | Davis and Venkatesh (2004) |
| | UB 2: I use mobile health service currently | |
| | UB 3: I spend a lot of time on mobile health service | |

*Table 1 Summary of Measurement Items*

## 4.2 Questionnaire Design and Data Collection

Items for questionnaire are developed from prior studies and modified or extended to fit the research context of our study. Each statement of the English questionnaire will be translated into Bengoli by researcher. The questionnaire will be divided into Part A and B. Part A contains the demographic information. Respondents will be asked information about their gender, age, marital status, educational qualifications, and mobile usage experience. Part B includes questionnaires for the different constructs presented in the proposed research model, using a 7-point Likert scale ranging from (1) "strongly disagree" to (7) "strongly agree. A pilot study will be conducted on selected participants in order to refine the questions and gain additional suggestions.

Potential participants will be sourced from residents living in two residential aged care facilities, called an old home in Bangladesh, 'Probin Hitoshi in Bangladesh", E-10; Probin Bhaban and Hospital; Agargaon; Sher-e- Banglanagar; Dhaka; Bangladesh and Old Rehabilitation Center, Bishia, Monipur, Gazipur, Bangladesh. The researchers will contact Director of those old homes seeking his/her permission to contact their residents for their participation in the study. About 300 participants will be recruited randomly. The structured questionnaire will be distributed among the randomly selected respondents (Fink, 2013). Individuals willing to participate will be given an information sheet with details about the project and their right to discontinue participation at any time. The participants will be given the option of answering the questions on paper themselves or recording their voice if unable to write and read. This research has been approved by the Human Research Ethics Committee at Southern Cross University, Australia with approval number is ECN-14-283.

## 4.3 Data Analysis

Partial Least Squares (PLS) method, a statistical analysis technique based on the Structural Equation Model (SEM), will be used to test the relationship among the constructs proposed in the research model. In PLS-SEM, sample size should be equal to 10 times greater than the largest number of indicators used to measure a single construct (Hair et al., 2013). Figure 1 shows that the proposed theoretical model consists of 9 constructs: PE, EE, SI, FC, HM, PV, HT, BI, and UB. As per measurement items (Table 1), total of 29 items for latent constructs in the proposed model are developed from prior studies and modified on the mHealth context in Bangladesh. Considering the proposed research model and the sample size recommended in PLS-SEM, 300 respondents will be sought in this study. The age range of potential participants will be 60+ as the study aiming to investigate factors influencing the adoption of mHealth by elderly.

We will assess the measurement model by examining the internal reliability, convergent and discriminant validity (Gotz et al., 2010). The internal reliability will be evaluated considering Cronbach's alpha and composite reliability. Convergent validity will be assessed by an average variance extracted (AVE) and items loading. In contrast, the discriminant validity will be assessed by the square root of the AVE and cross loading matrix (Hair et al., 2013). The structural model will be developed to identify the relationships among the constructs in the research model. Bootstrap method will be used to test the hypothesis to answer the research questions. In the first stage, the study will test the relationship between dependent and independent variables by path coefficient ($β$) and $t$-statistics. In the second stage, the study will discover the role of moderators. The demographic profile of respondents will also be presented.

# 5 CONCLUSIONS

## 5.1 Expected Outcome

The outcomes of this study will offer theoretical and practical guideline to the successful implementation and adaptation of mHealth services in Bangladesh as well as other developing nations. The proposed research will apply the UTAUT model to determine elderly behavioural intention to adopt mHealth applications. It is expected that the research findings will identify significant relationship between attitudinal constructs such as Performance Expectancy, Effort Expectancy, Social Influence, Facilitating Condition, Hedonic Motivation, Price Value, Habit and acceptance of mHealth in Bangladesh. The study is also expected to provide an overview of moderating role of gender and mobile phone uses experience and its influence on the adoption of mHealth technology and services.

## 5.2 Study Limitation

There are some limitations in the study due to surveying on age-specific participants and other factors. The proposed study intends to collect data only on the sample of elderly people which may raise concern about the generalizability of the findings. A further research, therefore, would extend the current study to younger and other aged group of citizens to uncover a more generalised view of the proposed model. Secondly, the nature of this study is cross-sectional. Therefore, the proposed study cannot confirm the causality and contingent effects of users' level of experience before and after the mHealth system and service adoption. Further study could use longitudinal data to unfold the causal relationship among variables over time. Another further study could also adopt both quantitative and qualitative approaches to unveil user's in-depth view and/or opinion on the issues.

## 5.3 Expected Benefits

**To participants:** The study offers respondents (elderly people in Bangladesh) an opportunity to reflect on their experiences about mHealth services in Bangladesh. Moreover, it is expected that the participants will gain a sense of satisfaction in knowing that their feedback on mHealth services is deemed important for future mHealth service design.

**To the broader community:** The empirical findings will provide insight into respondents experiences regarding mHealth services in Bangladesh and will seek to identify factors influencing the adoption of mHealth by elderly. With an increased knowledge of elderly perceptions of mHealth services, an expected benefit for the mHealth service providers and government is to understand the challenges/issues in regards to the design and implementation of successful mHealth services for its citizens, especially elderly community. The findings of the study could contribute to the planning and up-take of mHealth services of other developing countries with similar socio-economic circumstances.

**To increasing knowledge:** It is anticipated that this study will lead to increased knowledge of the use and adoption of mHealth technology. The study contributes to Information System (IS) research by providing a theoretical framework for mHealth use and acceptance, particularly by older community. The study will provide a practical guideline to the successful adoption of mHealth services in developing nations.

## Acknowledgements

This work was supported by seed research grant of Southern Cross School of Business (SCSB), Southern Cross University, under grant No. (31496). The authors, therefore, gratefully acknowledge the SCBS technical and financial support.